\pdfoutput=1
\documentclass[twoside,12pt,a4paper]{cmft}
\pagespan{1}{?}
\theoremstyle{remark}

\usepackage{amssymb}

\usepackage{hyperref}
\usepackage{caption}
\usepackage{subfig}

\usepackage{multirow}

\usepackage{bbding}

\usepackage{tikz}
\usetikzlibrary{shapes,arrows}
\usetikzlibrary{positioning}
\usepackage{pgfplots}

\newcounter{minutes}
\divide\time by 60
\newcounter{hours}
\multiply\time by 60
\addtocounter{minutes}{-\time}
\begin{document}

% Continuous Testing from Regression History into Smoke Testing Phase: An Industrial Case Study
\title[Continuous and Resource Managed Regression Testing: An Industrial Use Case]{Continuous and Resource Managed Regression Testing:  An Industrial Use Case}
%\title[Test Case Prioritisation and Selection Based on Codebase Changes and Regression History]{Test Case Prioritisation and Selection Based on \\ Codebase Changes and Regression History}
%\date{January 1, 2002}
\date{\today}

\author[T. Quach]{Tri Quach}
\email{tri.quach@siili.com}
\address{Siili Solutions,
         Porkkalankatu 24,
         00180,
         Finland}

%delete this block if only one author, copy if more than two authors
%\author[T. Quach]{Tri Quach${}^\textrm{{\tiny\EightFlowerPetal}}$}
\author[T. Oinonen]{Tommi Oinonen}
\email{tommi.oinonen@siili.com}
\address{Siili Solutions,
         Porkkalankatu 24,
         00180,
         Finland}

\author[A. Karjalainen]{Antti Karjalainen}
\email{antti.karjalainen@siili.com}
\address{Siili Solutions,
         Porkkalankatu 24,
         00180,
         Finland}

\keywords{functional testing, regression testing, test case prioritisation, industrial study}

\subjclass{}

% fill out if necessary or keep empty, acknowledgements go before bibliography
%\thanks{${}^\textrm{{\tiny\EightFlowerPetal}}$ The author was supported by a grant (MA2012n21) from the Magnus Ehrnrooth Foundation.}
\thanks{This research was supported by Business Finland.}
%\dedicatory{}

\begin{abstract}
Regression testing is an important part of quality control in both software and embedded products, where hardware is involved. It is also one of the most expensive and time consuming part of the product cycle. To improve the cost effectiveness of the development cycle and the regression testing, we use test case prioritisation and selection techniques to run more important test cases earlier in the testing process. 

In this paper, we consider a functional test case prioritisation with an access only to the version control of the codebase and regression history. Prioritisation is used to aid our test case selection, where we have chosen 5--25 (0.4\%--2.0\% of 1254) test cases to validate our method. The selection technique together with other prioritisation methods allows us to shape the current static, retest-all regression testing into a more resource managed regression testing framework. This framework will serve the agile way of working better and will allow us to allocate testing resources more wisely. This is a joint work with a large international Finnish company in an embedded industrial domain.

%In this paper, we consider a functional test case selection with a large international Finnish company in an embedded industrial domain. In this use case, we have access only to the version control of the codebase and regression history. The selection technique together with other prioritisation methods allows us to shape the current static, retest-all regression testing into a more resource managed regression testing framework. This framework will serve the agile way of working better and will allow us to allocate testing resources more wisely.

%By combining the selection technique with other ideas from other prioritisation methods, we will shape the current static regression testing into a more on-demand regression testing framework. This would serve the agile way of working better and allow us to allocate testing resources more wisely.

%We will combine the selection technique and ideas from other prioritisation methods to shape the static regression testing into a more dynamic regression testing framework, which would better serve agile way of working and better accommodate testing resources.

%Our goal is to shorten feedback loop to developers by bringing regression testing closer to smoke and sanity testing phases. 
%We also combine ideas from other prioritisation methods to shape static regression testing into more dynamic regression testing, which would better serve agile way of working and accommodate testing resources.

%by selecting only few functional test cases and run them several times during a day .
\end{abstract}

\maketitle

%\bigskip
%\begin{center}
%\texttt{FILE:~\jobname .tex, 2009-11-17,
%        printed: \number\year-\number\month-\number\day,
%        \thehours.\ifnum\theminutes<10{0}\fi\theminutes}
%\end{center}

\vspace*{-1.5cm}

\section{Introduction}
\label{sec:introduction}
In industries, regression testing is an important part of quality control and it is also used to ensure that the functionality of the system is not affected by modifications to the software. In practice, regression testing is in a constant change as there will be new test cases and old ones are modified or deprecated. Thus, widely used coverage based prioritisation methods are not optimal \cite{kaushik}. 

In agile software development with frequent release cycles, the system under test (SUT) is changing. Therefore, test selection techniques based on code change history and regression testing \cite{ekelund} are more effective than regression testing selection techniques that rely on static code analysis \cite{rothermel}. An in-depth analysis of different regression test selection techniques can be found in a systematic review article \cite{engstrom}, and a survey on regression testing minimisation, selection, and prioritisation methods is given in \cite{yoo-survey}. %Lastly, a systematic and extensive literature review on test case prioritisation techniques \cite{khati}. Most of the prioritisation techniques are based on coverage measure, and there are few studies based on combining codebase and regression history.

% Tästä voisi alkaa varsinainen intro omaan tekemiseen.
Regression testing is usually done by retesting all test cases at regular intervals. In our use case, running all the functional test cases takes more than one week. Thus, in practice, the whole regression test set is divided into daily and weekend regression subsets. The verdict of these regressions forms the baselines of the SUT. This testing practice is both expensive and time consuming \cite{runeson}, because, in the end, the majority of the functional test cases have passed through the entire collected regression history. On the other hand, even daily regression testing takes a long time for developers working in an agile environment, where changes to the software are made every day. Therefore, testing done during office hours should be more than just running build verification tests.
% and testing should accommodate the modified software.

In this paper, we propose a resource managed regression testing framework as an improvement to traditional retest-all regression testing. The framework is enabled by the algorithm proposed by Ekelund and Engstr\"om \cite{ekelund}, which we have validated in our client's systems. In our work, we have extended the original algorithm to take into account noise and memory handling, which Ekelund and Engstr\"om mentioned as downsides of their algorithm. Based on the extensions, we propose a way to utilise the method closer to the build verification testing phase by running a selection of few functional test cases several times during a day. 

Our use case study is only a part of the whole system in terms of regression history. In this particular subsystem, we have a total of 176 builds, 6720 modified files, and 1254 functional test cases in the regression history. The files can be divided into actual code and test files. However, in this study, we have not distinguished them from each other.  Furthermore, we have considered only functional test cases that have their verdict flipped at least once during the regression history. In this context, a flipped test case means that the verdict of the test case has been changed from passed to fail or vice versa as defined in \cite{ekelund}. Out of 176 builds we have 132 builds that have predictable test cases, i.e., test cases which have been flipped at least twice in the collected regression history.

The rest of the paper is organized as follows: Section \ref{sec:related_works} outlines related works. Our extension to the test selection algorithm, proposed in \cite{ekelund}, is described in Section \ref{sec:sensitivity-prioritisation}. Metrics and results of the algorithm are given in Section \ref{sec:metrics} and \ref{sec:results}, respectively. In Section \ref{sec:resource-reg}, we will discuss our resource managed regression testing framework, which combines sensitivity, history-based, and similarity-based prioritisation methods to improve the regression testing flow and to better accommodate testing resources. Finally, conclusions are given in Section \ref{sec:conclusions} along with future works.

\section{Related Works}
\label{sec:related_works}

In this section, we review different prioritisation and selection techniques related to our idea of shaping the regression testing. %We will explain how we are applying these method to achieve our vision. 
The methods below, with the exception of clustering methods, can be applied to situation with little to no prior data. In these kinds of situations the initial cost of applying the methods is low.

In agile software development one wants to select test cases based on code changes and regression testing \cite{ekelund} to catch the flipping test cases and to decrease the feedback loop time. These methods alone are not enough, because they do not catch test cases that keep on failing. Therefore, we need, e.g., a history-based test case prioritisation (HBTP) technique as well, where test cases are prioritised based on their failure rate in the regression history. Basically, if a given test case has failed then it will most likely fail again. One technique \cite{mantyla} gives weight to test cases depending on how many builds there have been since their last failure.

On the other hand, we want to execute test cases from different parts of the system in order to a have high system level coverage. This can be achieved by computing a string distant measure \cite{ali-mantyla,ledru} to measure how dissimilar test cases are. The dissimilarity prioritisation methods are useful, in particular, when test cases have always passed in the regression history. The selected test cases can be further prioritised with the distance measure as well, because similar test cases may test similar components of the system and thus detect the same fault.

Another history-based method is to cluster test cases based on regression history alone or on codebase changes. For example, one can build a co-change matrix of the modified files and compute a singular value decomposition to cluster the files. Then combining the clusters with the information on test cases, the method yields a list of prioritised test cases \cite{sherriff}. Using clustering methods, one may gain in-depth knowledge on how test cases behave, e.g., which test cases have passed and failed as a group even when that is not apparent by looking at the data alone. Downsides of clustering methods are, e.g., the required prior data and running the clustering algorithm in regular intervals to keep clusters up to date, which may become expensive in the long run.

%Empirically, if a given test case has failed then it will most likely fail again. Thus, the test case will be given a higher priority. These techniques belong to history-based test case prioritisation. 

%Another approaches are history-based test case prioritisation, where the test cases are prioritised based on their failure rate in the regression history \cite{mantyla}. Basically, if a given test case has failed often in recent history, then it would most likely to fail again, and, thus, it will be given a higher priority.

%There is a recent literature review article on different test ca

%Sherriff, Lake, and William \cite{sherriff} had an industrial case studies on test prioritisation based on a singular value decomposition, which was based on a co-change matrix of the modified files.

%and Osi{\'n}ski et al. \cite{osinski}.

\section{Sensitivity Matrix and Prioritisation}
\label{sec:sensitivity-prioritisation}

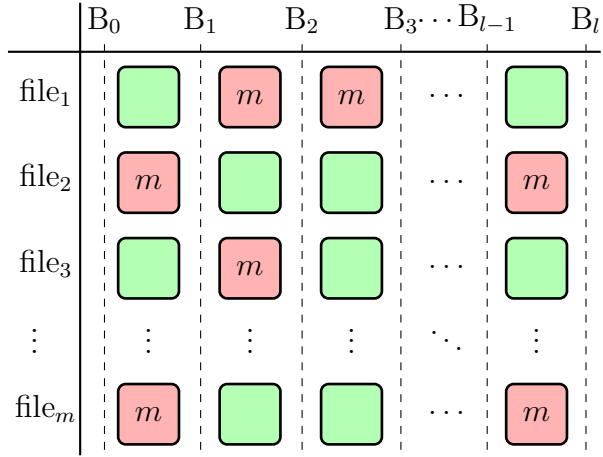
\begin{figure}[!ht]
\begin{center}
\begin{tikzpicture}[>=stealth]

\tikzstyle{block} = [draw, shape=rectangle,rounded corners=1mm, minimum height=0.8cm, minimum width=0.8cm, node distance = 0.3cm and 0.5cm, line width=1pt]
\tikzstyle{block1} = [draw, shape=rectangle,rounded corners=1mm, minimum height=0.8cm, minimum width=0.8cm, node distance = 0.3cm and 0.5cm, line width=1pt]
\tikzstyle{block2} = [draw, shape=rectangle,rounded corners=1mm, minimum height=0.8cm, minimum width=0.8cm, node distance = 0.3cm and 1.6cm, line width=1pt]
\tikzstyle{block4} = [draw, shape=rectangle,rounded corners=1mm, minimum height=0.8cm, minimum width=0.8cm, node distance = 1.1cm and 2.5cm, line width=1pt]

% tc1 builds
\node [block, fill=green!30] at (0,-1) (h11) {};
\node [block, fill=red!30, right =of h11] (h12) {$m$};
\node [block, fill=red!30, right =of h12] (h13) {$m$};
\node [block2, fill=green!30, right =of h13] (h1n) {};
%\node[right =of h13, anchor=south,midway] {$\ldots$};

\node [block, fill=red!30, below =of h11] (h21) {$m$};
\node [block, fill=green!30, right =of h21] (h22) {};
\node [block, fill=green!30, right =of h22] (h23) {};
\node [block2, fill=red!30, right =of h23] (h2n) {$m$};

\node [block, fill=green!30, below =of h21] (h31) {};
\node [block, fill=red!30, right =of h31] (h32) {$m$};
\node [block, fill=green!30, right =of h32] (h33) {};
\node [block2, fill=green!30, right =of h33] (h3n) {};

\node [block4, fill=red!30, below =of h31] (h41) {$m$};
\node [block, fill=green!30, right =of h41] (h42) {};
\node [block, fill=green!30, right =of h42] (h43) {};
\node [block2, fill=red!30, right =of h43] (h4n) {$m$};

% Lines
\draw [thick] (h11) ++(-1.85,0.6)  -- ++(7.8,0);
\draw [thick] (h11) ++(-0.9,1.25)  -- ++(0,-6);

% Title
%\draw (h11) ++(0.49,1.5) node[anchor=base] {Codebase Change History};

% Builds
\draw[thin, dashed] (h11) ++(-0.58,0.7) node[anchor=south] {$\textrm{B}_0$} ++(0,-0) -- ++(0,-5.48);
\draw[thin, dashed] (h12) ++(-0.65,0.7) node[anchor=south] {$\textrm{B}_1$} ++(0,-0) -- ++(0,-5.48);
\draw[thin, dashed] (h13) ++(-0.65,0.7) node[anchor=south] {$\textrm{B}_2$} ++(0,-0) -- ++(0,-5.48);
\draw[thin, dashed] (h13) ++(0.65,0.7) node[anchor=south] {$\textrm{B}_3$} ++(0,-0) -- ++(0,-5.48);

\draw[thin, dashed] (h1n) ++(-0.65,0.7) node[anchor=south] {$\textrm{B}_{l-1}$} ++(0,-0) -- ++(0,-5.48);
\draw[thin, dashed] (h1n) ++(0.65,0.7) node[anchor=south] {$\textrm{B}_{l}$} ++(0,-0) -- ++(0,-5.48);

% file
\draw (h11) ++(-1.35,-0.05) node[anchor=base] {$\textrm{file}_1$};
\draw (h21) ++(-1.35,-0.05) node[anchor=base] {$\textrm{file}_2$};
\draw (h31) ++(-1.35,-0.05) node[anchor=base] {$\textrm{file}_3$};
\draw (h41) ++(-1.35,-0.25) node[anchor=south] (texti) {$\textrm{file}_m$};

%  test case dots
\draw (h31) ++(-1.5,-1.12) node[anchor=base] {$\vdots$};
%\draw (h31)  ++(1.5,-1.12) node[anchor=base] {$\vdots$};
\draw (h31) ++(0,-1.12) node[anchor=base] {$\vdots$};
\draw (h32) ++(0,-1.12) node[anchor=base] {$\vdots$};
\draw (h33) ++(0,-1.12) node[anchor=base] {$\vdots$};
\draw (h3n) ++(0,-1.12) node[anchor=base] {$\vdots$};

\draw (h33) ++(1.25,-1.12) node[anchor=base] {$\ddots$};

% build dots
\draw (h13) ++(1.1,1) node[anchor=base] {$\ldots$};
\draw (h13)  ++(1.25,0) node[anchor=base] {$\ldots$};
\draw (h23) ++(1.25,0) node[anchor=base] {$\ldots$};
\draw (h33) ++(1.25,0) node[anchor=base] {$\ldots$};
\draw (h43) ++(1.25,0) node[anchor=base] {$\ldots$};
%\draw (h43) ++(0,-0.12) node[anchor=base] {$\vdots$};

\end{tikzpicture}

\caption{General structure of the codebase change history, where $m$-block means the corresponding files have been changed between two consecutive builds. } \label{fig: codebase-change-history}
\end{center}
\end{figure}

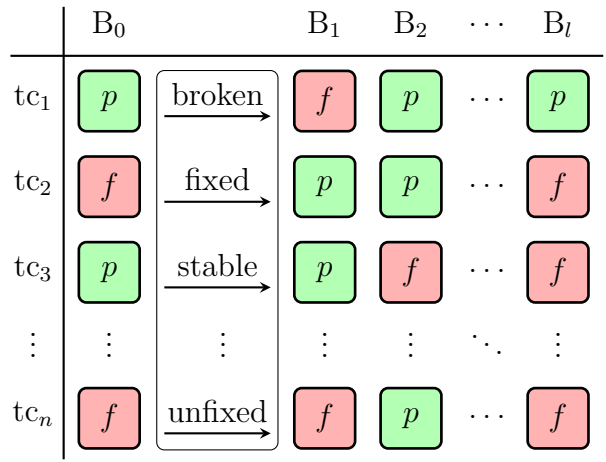
\begin{figure}[!ht]
\begin{center}
\begin{tikzpicture}[>=stealth]
\tikzstyle{block} = [draw, shape=rectangle,rounded corners=1mm, minimum height=0.8cm, minimum width=0.8cm, node distance = 0.3cm and 2cm, line width=1pt]
\tikzstyle{block1} = [draw, shape=rectangle,rounded corners=1mm, minimum height=0.8cm, minimum width=0.8cm, node distance = 0.3cm and 0.3cm, line width=1pt]
\tikzstyle{block2} = [draw, shape=rectangle,rounded corners=1mm, minimum height=0.8cm, minimum width=0.8cm, node distance = 0.3cm and 1.1cm, line width=1pt]
\tikzstyle{block3} = [draw, shape=rectangle,rounded corners=1mm, minimum height=0.8cm, minimum width=0.8cm, node distance = 0.3cm and 1.1cm, line width=1pt]
\tikzstyle{block4} = [draw, shape=rectangle,rounded corners=1mm, minimum height=0.8cm, minimum width=0.8cm, node distance = 1.1cm and 2cm, line width=1pt]

% tc1 builds
\node [block, fill=green!30] at (0,-1) (h11) {$p$};
\node [block, fill=red!30, right =of h11] (h12) {$f$};
\node [block1, fill=green!30, right =of h12] (h13) {$p$};
\node [block2, fill=green!30, right =of h13] (h1n) {$p$};
%\node[right =of h13, anchor=south,midway] {$\ldots$};

\node [block, fill=red!30, below =of h11] (h21) {$f$};
\node [block, fill=green!30, right =of h21] (h22) {$p$};
\node [block1, fill=green!30, right =of h22] (h23) {$p$};
\node [block2, fill=red!30, right =of h23] (h2n) {$f$};

\node [block, fill=green!30, below =of h21] (h31) {$p$};
\node [block, fill=green!30, right =of h31] (h32) {$p$};
\node [block1, fill=red!30, right =of h32] (h33) {$f$};
\node [block2, fill=red!30, right =of h33] (h3n) {$f$};

\node [block4, fill=red!30, below =of h31] (h41) {$f$};
\node [block, fill=red!30, right =of h41] (h42) {$f$};
\node [block1, fill=green!30, right =of h42] (h43) {$p$};
\node [block2, fill=red!30, right =of h43] (h4n) {$f$};

\draw[thin,rounded corners=1mm] (h11) ++(0.625cm,0.405cm) -- ++(0,-5.03cm) -- ++(1.595cm,0) -- ++(0,5.03cm) -- cycle;
\draw[->,thick] (h11)  ++(0.725,-0.2) -- ++(1.4,0) node[anchor=south,midway] {broken};
\draw[->,thick] (h21)  ++(0.725,-0.2) -- ++(1.4,0) node[anchor=south,midway] {fixed};
\draw[->,thick] (h31)  ++(0.725,-0.2) -- ++(1.4,0) node[anchor=south,midway] {stable};
\draw[->,thick] (h41)  ++(0.725,-0.2) -- ++(1.4,0) node[anchor=south,midway] {unfixed};

% Lines
\draw [thick] (h11) ++(-1.3,0.6)  -- ++(7.8,0);
\draw [thick] (h11) ++(-0.6,1.25)  -- ++(0,-6);

% Title
%\draw (h11) ++(0.35,1.5) node[anchor=base] {Regression History};

% Builds
\draw (h11) ++(0,0.7) node[anchor=south] {$\textrm{B}_0$};
\draw (h12) ++(0,0.7) node[anchor=south] {$\textrm{B}_1$};
\draw (h13) ++(0,0.7) node[anchor=south] {$\textrm{B}_2$};
\draw (h1n) ++(0,0.7) node[anchor=south] {$\textrm{B}_l$};

% Test case
\draw (h11) ++(-1,-0.05) node[anchor=base] {$\textrm{tc}_1$};
\draw (h21) ++(-1,-0.05) node[anchor=base] {$\textrm{tc}_2$};
\draw (h31) ++(-1,-0.05) node[anchor=base] {$\textrm{tc}_3$};
\draw (h41) ++(-1,-0.25) node[anchor=south] {$\textrm{tc}_n$};

%  test case dots
\draw (h31) ++(-1,-1.12) node[anchor=base] {$\vdots$};
\draw (h31)  ++(1.5,-1.12) node[anchor=base] {$\vdots$};
\draw (h31) ++(0,-1.12) node[anchor=base] {$\vdots$};
\draw (h32) ++(0,-1.12) node[anchor=base] {$\vdots$};
\draw (h33) ++(0,-1.12) node[anchor=base] {$\vdots$};
\draw (h3n) ++(0,-1.12) node[anchor=base] {$\vdots$};

\draw (h33) ++(1,-1.12) node[anchor=base] {$\ddots$};

% build dots
\draw (h13) ++(1,1) node[anchor=base] {$\ldots$};
\draw (h13)  ++(1,0) node[anchor=base] {$\ldots$};
\draw (h23) ++(1,0) node[anchor=base] {$\ldots$};
\draw (h33) ++(1,0) node[anchor=base] {$\ldots$};
\draw (h43) ++(1,0) node[anchor=base] {$\ldots$};
%\draw (h43) ++(0,-0.12) node[anchor=base] {$\vdots$};

\end{tikzpicture}
\caption{General structure of the test regression history with labels for four different possible transitions based on the initial verdict of  given test cases.} \label{fig: regression-history}
\end{center}
\end{figure}

The prioritisation method is based on collecting lists of the modified files from the version control and verdicts of the regression history, see Figure \ref{fig: codebase-change-history} and Figure \ref{fig: regression-history}. In industrial scale, both information are usually available, but the link between a specific regression history run and the SUT may not be available. Without the knowledge of the link, we cannot determine the modified files from the version control between two builds and their regression verdicts.

%\subsection{Sensitivity Matrix}
%\label{sec:sensitivity-matrix}

For each build, we construct a sensitivity matrix $B_k$ as follows
\begin{equation}
b_{ij}^k = \left\{ \begin{split} 
1/d(|\textrm{fc}_k|), & \textrm{ file}_i \in \textrm{fc}_k \land \textrm{tc}_j \in \textrm{flipped}_k, \\
0, & \textrm{ else},\end{split}\right. \label{eqn:sensitivity}
\end{equation}
where function $d$ is an increasing function with respect to $|\textrm{fc}_k|$, $\textrm{fc}_k$ is a set of files that have been modified between two consecutive builds, and $\textrm{flipped}_k$ is the set of flipped test cases. The function $d$ acts as a confidence on a direct correlation between files and test cases. In this particular use case, we have chosen $d = |\textrm{fc}_k|$. 

%It should be noted that the function $d$ must be chosen by keeping in mind the 

%\subsection{Pseudo Coverage}
We define the sensitivity prioritisation matrix as follows
\begin{equation}
\left\{ \begin{split}
\tilde{B}_0 & = B_0  = 0, \\
\tilde{B}_k & = \alpha B_k + (1-\alpha) \tilde{B}_{k-1},\quad k \geq 1, \, \alpha \in [0,1].
\end{split} \right.  \label{eqn:ema-prio}
\end{equation} 
Equation \eqref{eqn:ema-prio} is called as an exponential moving average (EMA), where the coefficient $\alpha$ is a weighting factor for the new observation and a decaying factor for the older observation as suggested by Kim and Porter in \cite{kim}. To prioritise the test cases in the build $k$, we slice the sensitivity prioritisation matrix $\tilde{B}_{k-1} $ by taking the rows corresponding to the modified files in $\textrm{fc}_k$ and we call it $\tilde{B}_{k-1}^{\textrm{slice}}$. The column sum of $\tilde{B}_{k-1}^{\textrm{slice}}$ gives us the sensitivity of test cases over the file changes. In other words, a higher sum means that the test case is more likely to flip its verdict. To prioritise test cases one can use the maximum element of the columns as well instead of the column sum. It should be noted that \eqref{eqn:ema-prio} with $d \equiv 1$  and without the coefficients $\alpha$ and $1-\alpha$ is exactly the method proposed in \cite{ekelund}.

\subsection{Heat Map}
The sensitivity prioritisation matrix can be used in analysing the behaviour of the system by computing the heat map. One may reveal the flakiness of a test case by looking at the heat map. Let us assume that a test case tests a certain feature of the system and that the feature is affected by a handful of files. If a test case has sensitivity towards the majority of the files and the sensitivity itself is relatively small, then the test case itself is sensitive to modifications. Some of these sensivity test cases are flaky or bad.

For another example, we assume that a HBTP method flags test cases that should be run based on their recent failure rate. We can look up the corresponding column of the test cases from $\tilde{B}$ and the rows (files) of the largest elements. These files have had the greatest impact on flipping the test case during the regression history and developers should take a closer look at them.

\section{Metrics}
\label{sec:metrics}

To measure our techniques we use the framework and definitions given in \cite{rothermel}. In this sense, we measure the predictability of the test cases. A test case is defined as predictable in the build $B_k$ if it has its verdict flipped between the builds $B_{k-1}$ and $B_k$, and if it has flipped at least once in the history before. A natural choice for measurement is to compute precision and recall, which are defined as follows
\begin{equation} 
\textrm{precision} = \frac{|\textrm{selected} \cap \textrm{predictable}|}{|\textrm{selected}|}, \label{eqn:precision}
\end{equation}
and
\begin{equation}
\textrm{recall} = \frac{|\textrm{selected} \cap \textrm{predictable}|}{|\textrm{predictable}|}, \label{eqn:recall}
\end{equation}
where the set of selected test cases is the top prioritised test cases from the sliced sensitivity prioritisation matrix $\tilde{B}_{k-1}^{\textrm{slice}}$. The range for both, precision and recall, is $[0,1]$ and they can be combined into a single metric 
\begin{equation}
\textrm{F-measure} = 2 \frac{\textrm{precision} \cdot \textrm{recall}}{\textrm{precision}+\textrm{recall}}, \label{eqn:f-measure}
\end{equation}
which can be used to compare different techniques. 

Let us assume $\{\textrm{predictable}\} \subset \{\textrm{selected}\}$, in which case selecting more test cases decreases precision, while the recall remains unchanged and equals $1$. Thus, the F-measure will perform worse than if one had selected fewer test cases. 
%On the other way around, the F-measure performs either better or worse.
%the set of predictable test cases is a subset of the set of selected test cases, 

%In practise, due to our goal to test more often during a day by selecting only a several test cases, the F-measure is not optimal. This is because of the definition of the recall and precision. In our case, we choose to select on few functional test cases and if the set of valid test cases are way larger than set selected functional test cases, then the recall will be small. This is 

%If we wish to obtain the maximum value or, in general, have the whole range to be attainable, we need to select as many test cases as there are valid test cases. Therefore, in this study, we are using the maximum of the recall and precision to measure our techniques. 

\section{Results}
\label{sec:results}
We compared our sensitivity prioritisation method to the method given in \cite{ekelund} and to selecting random test cases by computing the averages of precision, recall, and the F-measure for different sizes of selected test cases. We also counted how many times the methods returned zero results, i.e., the builds where we have $\{\textrm{selected}\} \cap \{\textrm{predictable}\} = \varnothing$. In this study, we chose to select 5--25 (0.4\%--2\%) test cases, which is in line with our goal to run a small number of selected test cases several times a day.

From 132 builds, we had 41 (31\%) builds with 5 or fewer predictable test cases and 76 (58\%) builds with more than 25 predictable test cases. The hyperparameter $\alpha$ is chosen by minimising the zero results over the regression history. In our use case, the optimal value turned out to be $\alpha = 0.80$ for our EMA sensitivity prioritisation method. For our randomised test case selection, we chose to select test cases randomly from all the 1254 test cases. For each selection size, we took an average of over 100 runs. Note that, a short term improvement to the randomised selection can be achieved by selecting test cases that have been flipped up to the corresponding build. 

The results from our algorithm and \cite{ekelund} are collected into Table \ref{table:results} along with the minimal, maximal, and average improvements we have achieved. The number of times the algorithm gives zero correct predictable test cases depends heavily on the number of selected test cases. Thus, the first interesting point in the results is how randomised selection yielded surprisingly a comparable result on percentage of zero cases to the algorithm given in \cite{ekelund}. As for the rest of the metrics, randomisation does a poor job compared to \cite{ekelund} and our method. On the other hand, our version of the algorithm gave on average 35\% zero result on predictable test cases, which is 25\% fewer zero results compared to \cite{ekelund}. The result is a significant improvement to the amount of selected test cases considered. 

\begin{figure}[!ht]
\begin{center}
\begin{tikzpicture} 
\begin{axis}[ width=8.9cm, height=7cm, 
	xlabel=Number of Selected Test Cases, ylabel= Recall,
    y tick label style={
        /pgf/number format/.cd,
            fixed,
            fixed zerofill,
            precision=2,
        /tikz/.cd
    },
	ytick={0,0.05, 0.10, 0.15, 0.20},
	legend style={at={(0.38,0.17)},anchor=west,draw=black,fill=white,align=left}
	]
\addplot [color=red,mark=*, smooth] coordinates {
       (5, 0.121105)
       (6, 0.127854)
       (7, 0.134306)
       (8, 0.140180)
       (9, 0.143833)
       (10, 0.148033)
       (11, 0.152139)
       (12, 0.161943) 
       (13, 0.164199)
       (14, 0.169961)
       (15, 0.172076)
       (16, 0.174061)
       (17, 0.175648)
       (18, 0.178838)
       (19, 0.180448)
       (20, 0.192319)
       (21, 0.194687)
       (22, 0.195989)
       (23, 0.197397)
       (24, 0.198858)
       (25, 0.200313)
};

\addplot[color=blue,mark=diamond, smooth, dashed, mark options={solid}, thick] coordinates {
       (5, 0.047043)
       (6, 0.056453)
       (7, 0.058229)
       (8, 0.065306)
       (9, 0.068963)
       (10, 0.076310)
       (11, 0.081210)
       (12, 0.082238)
       (13, 0.083268)
       (14, 0.088239)
       (15, 0.089147) 
       (16, 0.090127)
       (17, 0.090916)
       (18, 0.096009)
       (19, 0.100940)
       (20, 0.103854)
       (21, 0.106419)
       (22, 0.107413)
       (23, 0.108713)
       (24, 0.109825)
       (25, 0.112864)
};

\addplot[color=green,mark=o, smooth, mark options={solid}, thick] coordinates {
       (5, 0.003800)
       (6, 0.004457)
       (7, 0.005946)
       (8, 0.006230)
       (9, 0.007028)
       (10, 0.008158)
       (11, 0.008530)
       (12, 0.009200)
       (13, 0.010568)
       (14, 0.011109)
       (15, 0.012210)
       (16, 0.012146)
       (17, 0.013663)
       (18, 0.014628)
       (19, 0.015846)
       (20, 0.015530)
       (21, 0.016512)
       (22, 0.017637)
       (23, 0.017969)
       (24, 0.018288)
       (25, 0.019908)
};

%\addplot +[mark=none] coordinates {(5, 0.4) (30, 0.4)};

\legend{EMA $\alpha=0.8$, Method \cite{ekelund}, Randomised}
\end{axis}
\end{tikzpicture}
%\vspace*{-0.50cm}
\caption{Average recall for each number of selected test cases.} \label{fig:recall}
\end{center}
\end{figure}
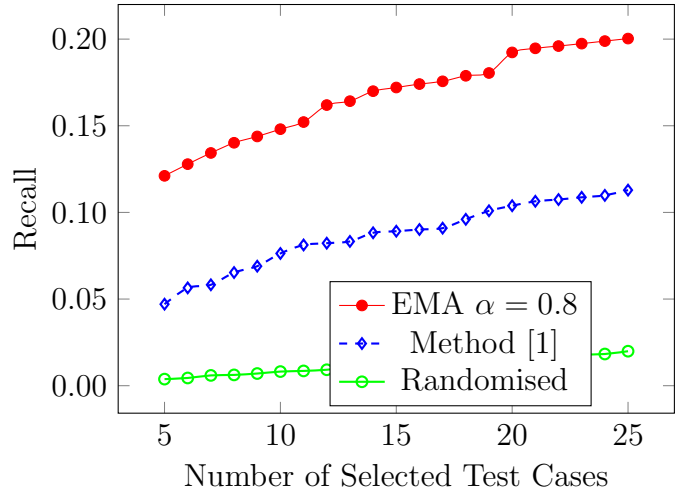

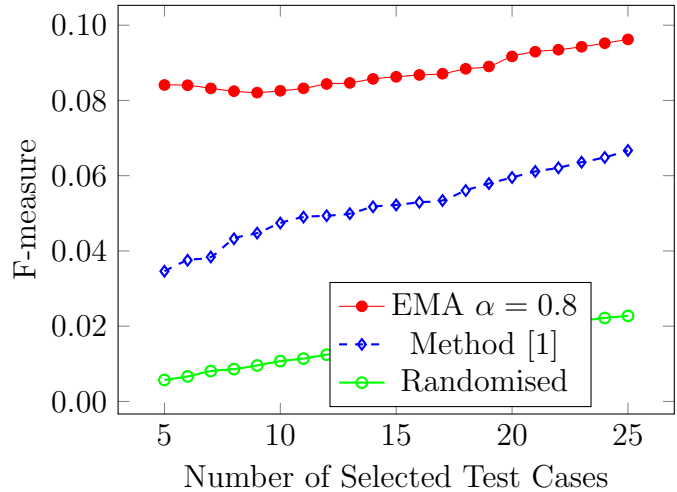
\begin{figure}[!ht]
\begin{center}
\begin{tikzpicture} 
\begin{axis}[ width=8.9cm, height=7cm, 
	xlabel=Number of Selected Test Cases, ylabel= F-measure,
    y tick label style={
        /pgf/number format/.cd,
            fixed,
            fixed zerofill,
            precision=2,
        /tikz/.cd
    },
	legend style={at={(0.38,0.17)},anchor=west,draw=black,fill=white,align=left}
	]
\addplot [color=red,mark=*, smooth] coordinates {
       (5, 0.084108)
       (6, 0.084043)
       (7, 0.083202)
       (8, 0.082465)
       (9, 0.082081)
       (10, 0.082568)
       (11, 0.083215)
       (12, 0.084373)
       (13, 0.084660)
       (14, 0.085730)
       (15, 0.086277)
       (16, 0.086774)
       (17, 0.087100)
       (18, 0.088403)
       (19, 0.089031)
       (20, 0.091694)
       (21, 0.092949)
       (22, 0.093477)
       (23, 0.094257)
       (24, 0.095212)
       (25, 0.096223)
};

\addplot[color=blue,mark=diamond, smooth, dashed, mark options={solid}, thick] coordinates {
       (5, 0.034606)
       (6, 0.037525)
       (7, 0.038386)
       (8, 0.043238)
       (9, 0.044746)
       (10, 0.047405)
       (11, 0.048990)
       (12, 0.049359)
       (13, 0.049907)
       (14, 0.051718)
       (15, 0.052219)
       (16, 0.052929)
       (17, 0.053413)
       (18, 0.056051)
       (19, 0.057908)
       (20, 0.059507)
       (21, 0.061119)
       (22, 0.062092)
       (23, 0.063581)
       (24, 0.064844)
       (25, 0.066708)
};

\addplot[color=green,mark=o, smooth, mark options={solid}, thick] coordinates {
       (5, 0.005706)
       (6, 0.006643)
       (7, 0.008063)
       (8, 0.008591)
       (9, 0.009561)
       (10, 0.010708)
       (11, 0.011383)
       (12, 0.012432)
       (13, 0.013289)
       (14, 0.014258)
       (15, 0.015009)
       (16, 0.015697)
       (17, 0.016649)
       (18, 0.017447)
       (19, 0.018290)
       (20, 0.019114)
       (21, 0.019793)
       (22, 0.020665)
       (23, 0.021456)
       (24, 0.022180)
       (25, 0.022745)
};

%\addplot +[mark=none] coordinates {(5, 0.4) (30, 0.4)};

\legend{EMA $\alpha=0.8$, Method \cite{ekelund}, Randomised}
\end{axis}
\end{tikzpicture}
%\vspace*{-0.50cm}
\caption{Average F-measure for each number of selected test cases.} \label{fig:f-measure}
\end{center}
\end{figure}

Our method shows a minor improvement in precision compared to \cite{ekelund}, but both results seem to be capped. The reason for that may have to do with the number of builds with fewer predictable test cases than the selected test cases as well as with the definition of precision. In recall metric, see Fig. \ref{fig:recall}, our algorithm shows on an average 97\% better result than \cite{ekelund}, which is a significant improvement as well. The result in recall is carried to the F-measure, see Fig. \ref{fig:f-measure}, as well.

%\begin{table}[h]
%\caption{Average results from our algorithm and the algorithm from \cite{ekelund} along with minimum, maximum and average improvement.
%  }\label{table:results}
%\begin{tabular}{c|c|c|ccc}
% &  EMA, & Ekelund, & & Improvement  \\
% & $\alpha=0.80$ & Engstr\"om \cite{ekelund} & min & max & average \\
%\hline
%Percentage of $\varnothing$ & 35\% & 47\% & -26\% & \\
%Precision & 0.36 & 0.13 & 175\% & \\
%Recall & 0.168 & 0.012 & 1310\%  & \\
%F-measure & 0.089 & 0.015 & 493\% &
%\end{tabular}
%\end{table}

\begin{table}[h]
\caption{Average results over 5--25 selected test cases for our algorithm and the algorithm from \cite{ekelund} along with minimum, maximum, and average improvement.
  }\label{table:results}
\begin{center}
\begin{tabular}{c|r|r|rrr|c}
    \multicolumn{1}{c}{} & \multicolumn{1}{|c|}{EMA}
      & \multicolumn{1}{|c|}{Method} & \multicolumn{3}{c}{Improvement} & \multicolumn{1}{|c}{Fig-}  \\
%    \cline{4-6}
    & $\alpha = 0.80$ & \cite{ekelund} & min & max & avg & ure \\ 
    \hline
Pct of $\varnothing$ & 35\% & 47\% & -22\% & -30\% & -25\% & \ref{fig:noz}  \\
Precision & 0.36 & 0.34 & 0.5\% & 8.7\% & 5.4\% & \ref{fig:precision}  \\
Recall & 0.168 & 0.089 & 77\% & 160\% & 97\% & \ref{fig:recall} \\
F-measure & 0.089 & 0.052 & 44\% & 140\% & 72\% & \ref{fig:f-measure} 
\end{tabular}
\end{center}
\end{table}

%[n_largest, noz, p, r, M , F, pnz, rnz, Mnz, Fnz]
%
%salabs
%[15.        0.3495670995454545  0.35640186  0.16781843  0.46785424  0.08751629
%  0.54980733  0.25678867  0.71968748  0.13464238]
%
%ekelund
%[15.000000 0.4693362196969697   0.338385 0.086833 0.394877 0.052202 0.645793 0.161673
% 0.750663 0.097642]
%
%versus improvement
%min
%[1.        , 0.7027027 , 1.0053052 , 1.77481748, 1.10835878,
%       1.44245068, 0.73822137, 1.48637916, 0.81438193, 1.25012844]
%
%max
%[1.        , 0.77777778, 1.08704914, 2.57434687, 1.23901438,
%       2.43044559, 0.93866953, 1.92218204, 1.07207196, 1.81471583]
%
%average
%[1.         0.74589411 1.05428462 1.97187836 1.1847568  1.71811133
% 0.85914451 1.59885519 0.96603602 1.38978192]

%\begin{figure}
%\begin{center}
%  \subfloat[Number of zero cases. Lower is better.]{\label{fig:noz}\includegraphics[width=0.45\textwidth]{SUM_Salabs_vs_Ekelund_vs_Random_id_8_alpha_80_number_of_zeros}}\quad
%  \subfloat[Precision]{\label{fig:precision}\includegraphics[width=0.45\textwidth]{SUM_Salabs_vs_Ekelund_vs_Random_id_8_alpha_80_precision}} \\
%  \subfloat[Recall]{\label{fig:recall}\includegraphics[width=0.45\textwidth]{SUM_Salabs_vs_Ekelund_vs_Random_id_8_alpha_80_recall}}\quad
%    \subfloat[F-measure]{\label{fig:f-measure}\includegraphics[width=0.45\textwidth]{SUM_Salabs_vs_Ekelund_vs_Random_id_8_alpha_80_F_measure}}
%  \caption{Metrics of our use case study.}
%\label{fig:results}
%\end{center}
%\end{figure}

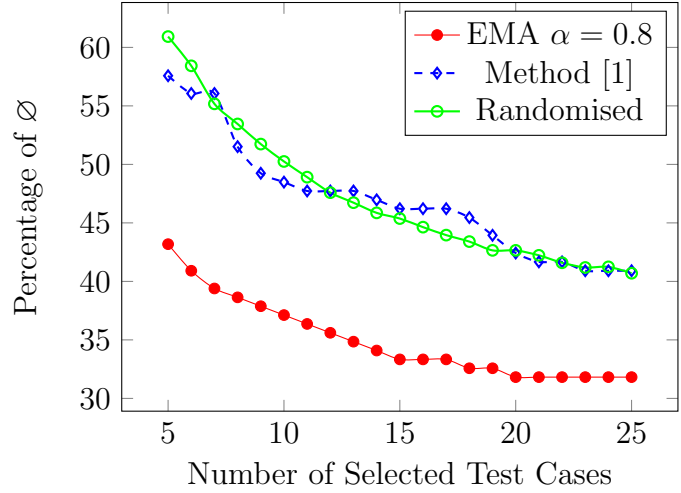
\begin{figure}[!ht]
\begin{center}
\begin{tikzpicture} 
\begin{axis}[ width=8.9cm, height=7cm, 
	xlabel=Number of Selected Test Cases, ylabel= Percentage of $\varnothing$,
	ytick={30,35,40,45,50,55,60}]
\addplot[color=red,mark=*, smooth] coordinates {
       (5, 57/132*100)
       (6, 54/132*100)
       (7, 52/132*100)
       (8, 51/132*100)
       (9, 50/132*100)
       (10, 49/132*100)
       (11, 48/132*100)
       (12, 47/132*100)
       (13, 46/132*100)
       (14, 45/132*100)
       (15, 44/132*100)
       (16, 44/132*100)
       (17, 44/132*100)
       (18, 43/132*100)
       (19, 43/132*100)
       (20, 42/132*100)
       (21, 42/132*100)
       (22, 42/132*100)
       (23, 42/132*100)
       (24, 42/132*100)
       (25, 42/132*100)
};

\addplot[color=blue,mark=diamond, smooth, dashed, mark options={solid}, thick] coordinates {
       (5, 76/132*100)
       (6, 74/132*100)
       (7, 74/132*100)
       (8, 68/132*100)
       (9, 65/132*100)
       (10, 64/132*100)
       (11, 63/132*100)
       (12, 63/132*100)
       (13, 63/132*100)
       (14, 62/132*100)
       (15, 61/132*100)
       (16, 61/132*100)
       (17, 61/132*100)
       (18, 60/132*100)
       (19, 58/132*100)
       (20, 56/132*100)
       (21, 55/132*100)
       (22, 55/132*100)
       (23, 54/132*100)
       (24, 54/132*100)
       (25, 54/132*100)
};

\addplot[color=green,mark=o, smooth, mark options={solid}, thick] coordinates {
       (5, 80.42/132*100)
       (6, 77.12/132*100)
       (7, 72.83/132*100)
       (8, 70.56/132*100)
       (9, 68.29/132*100)
       (10, 66.33/132*100)
       (11, 64.56/132*100)
       (12, 62.81/132*100)
       (13, 61.68/132*100)
       (14, 60.53/132*100)
       (15, 59.89/132*100)
       (16, 58.92/132*100)
       (17, 58.02/132*100)
       (18, 57.29/132*100)
       (19, 56.30/132*100)
       (20, 56.30/132*100)
       (21, 55.74/132*100)
       (22, 54.90/132*100)
       (23, 54.37/132*100)
       (24, 54.44/132*100)
       (25, 53.72/132*100)
};

%\addplot +[mark=none] coordinates {(5, 0.4) (30, 0.4)};

\legend{EMA $\alpha=0.8$, Method \cite{ekelund}, Randomised}
\end{axis}
\end{tikzpicture}
%\vspace*{-0.50cm}
\caption{Percentage of builds with $\{\textrm{selected}\} \cap \{\textrm{predictable}\} = \varnothing$.} \label{fig:noz}
\end{center}
\end{figure}

%%%%%%%%%%%%%%%%%%%%%%%%%%%%%%%%%%%%%%%%%%%%%%%%%%%%%%%%%%%%%%%
%%%%%%%%%%%%%%%%%%%%%%%%%%%%%%%%%%%%%%%%%%%%%%%%%%%%%%%%%%%%%%%

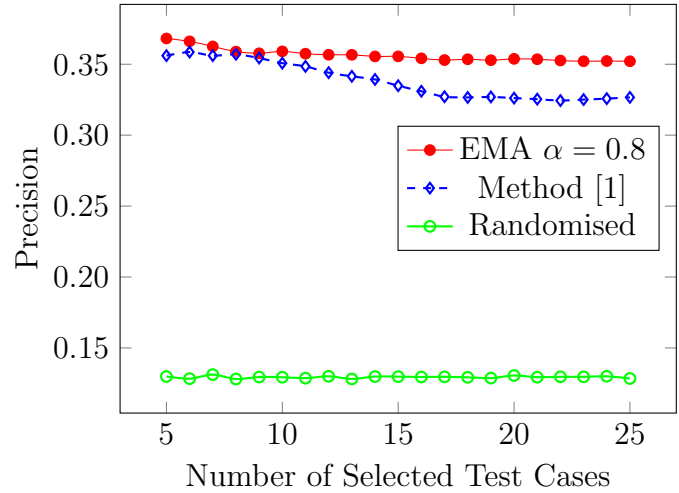
\begin{figure}[!ht]
\begin{center}
\begin{tikzpicture} 
\begin{axis}[ width=8.9cm, height=7cm, 
	xlabel=Number of Selected Test Cases, ylabel= Precision,
	ytick={0.15, 0.20, 0.25, 0.30, 0.35},
	    y tick label style={
        /pgf/number format/.cd,
            fixed,
            fixed zerofill,
            precision=2,
        /tikz/.cd
    },
	legend style={at={(0.97,0.55)},anchor=east,draw=black,fill=white,align=left}
	]
\addplot[color=red,mark=*, smooth] coordinates {
       (5, 0.368182)
       (6, 0.366162)
       (7, 0.362554)
       (8, 0.358902)
       (9, 0.357744)
       (10, 0.359091)
       (11, 0.357438)
       (12, 0.356692)
       (13, 0.356643)
       (14, 0.355519)
       (15, 0.355556)
       (16, 0.354167)
       (17, 0.352941)
       (18, 0.353535)
       (19, 0.352871)
       (20, 0.353788)
       (21, 0.353535)
       (22, 0.352617)
       (23, 0.352108)
       (24, 0.352273)
       (25, 0.352121)
};

\addplot[color=blue,mark=diamond, smooth, dashed, mark options={solid}, thick] coordinates {
       (5, 0.356061)
       (6, 0.358586)
       (7, 0.356061)
       (8, 0.357008)
       (9, 0.354377)
       (10, 0.350758)
       (11, 0.348485)
       (12, 0.344066)
       (13, 0.341492)
       (14, 0.339286)
       (15, 0.334848)
       (16, 0.330966)
       (17, 0.327094)
       (18, 0.326599)
       (19, 0.326954)
       (20, 0.326136)
       (21, 0.325397)
       (22, 0.324380)
       (23, 0.325099)
       (24, 0.325758)
       (25, 0.326667)
};

\addplot[color=green,mark=o, smooth, mark options={solid}, thick] coordinates {
       (5, 0.129864)
       (6, 0.128346)
       (7, 0.131190)
       (8, 0.128097)
       (9, 0.129478)
       (10, 0.129356)
       (11, 0.128705)
       (12, 0.130025)
       (13, 0.128153)
       (14, 0.129832)
       (15, 0.129773)
       (16, 0.129522)
       (17, 0.129639)
       (18, 0.129280)
       (19, 0.128736)
       (20, 0.130568)
       (21, 0.129383)
       (22, 0.129652)
       (23, 0.129615)
       (24, 0.130028)
       (25, 0.128461)
};

%\addplot +[mark=none] coordinates {(5, 0.4) (30, 0.4)};

\legend{EMA $\alpha=0.8$, Method \cite{ekelund}, Randomised}
\end{axis}
\end{tikzpicture}
%\vspace*{-0.50cm}
\caption{Average precision for each number of selected test cases.} \label{fig:precision}
\end{center}
\end{figure}

%%%%%%%%%%%%%%%%%%%%%%%%%%%%%%%%%%%%%%%%%%%%%%%%%%%%%%%%%%%%%%%
%%%%%%%%%%%%%%%%%%%%%%%%%%%%%%%%%%%%%%%%%%%%%%%%%%%%%%%%%%%%%%%

%%%%%%%%%%%%%%%%%%%%%%%%%%%%%%%%%%%%%%%%%%%%%%%%%%%%%%%%%%%%%%%
%%%%%%%%%%%%%%%%%%%%%%%%%%%%%%%%%%%%%%%%%%%%%%%%%%%%%%%%%%%%%%%

\section{Resource Managed Regression Testing}
\label{sec:resource-reg}

Our vision is to move from retest-all regression testing to an on-demand resource managed regression testing framework, where we allocate testing resources based on, e.g., the time of the day, the software modification, test history, and similarity of the test cases themselves. The framework, in general, contains the following steps:
\begin{itemize}
\item Select test cases based on code changes using \eqref{eqn:ema-prio}.
\item Prioritise test cases that fail regularly based on their historical verdicts. 
\item Run more stable test cases based on their dissimilarity measures.
\end{itemize}
During office hours, the first two bullet points are looped to gain faster feedback on features under development. More stable test cases are run after office hours similarly to daily regression. Instead of running static regression, we prioritise test cases based on their similarity measure and select dissimilar test cases to gain a higher system level coverage.

In order to update the sensitivity prioritisation matrix using this workflow, we need to keep track of modified files for each test case since the last time it has been executed. The update to the sensitivity prioritisation matrix will occur when the test cases are run the next time and the update is done test case wise. The development of particular features is more likely to flip certain test cases, and a rapid execution of these test cases may cause bias. Each time the verdict of the test case does not flip, the update procedure will decrease the sensitivity values exponentially and will thus reduce the bias.

Situations where some test cases have not been executed for a long time cause real concerns with regards to prioritising and executing more stable test cases. These concerns can be addressed with high level test strategy, which can be, e.g., round robin or minimisation of a cost function. In a round robin strategy, one can give a specific time frame in which all the stable test cases must be executed at least once. 

A cost function strategy may look as follows. Let $s_i$ be the number of days since the $i$th test case has been last executed. Then we define a cost function as
\[
\textrm{cost} = \sum_i s_i^2,
\]
which we want to minimise. The cost function will add another constraint to the prioritisation of the stable test cases. 

Lastly, by controlling test strategies, we can accommodate upcoming releases by allocating more resources to test the release candidate or any part of the system, that needs more focus and resources.

%propose a fuzzy coverage prioritisation method as a framework to improve regression testing. The idea 
%
%
%Selecting few test cases based on code modifications allows us to run 
%
%Prioritising test cases based on code modifications allows us to select few test cases to be run in  
%
%can be used to decrease feedback time to developers

\section{Conclusions} 
\label{sec:conclusions}
We have demonstrated a viable algorithm to select a small number of functional test cases with minimal prior data. These selected test cases can be run closer to the build verification testing phase to give developers faster feedbacks. Our algorithm also shows a significant improvement compared to the original algorithm given in \cite{ekelund}. The usage and results of our algorithm suggested that it can be combined with other prioritisation methods to shape the regression testing into a more resource managed regression testing.

However, there are improvements to be made to our algorithm and ideas. One may consider modified functions instead of modified files to have a better correlation between the functions and test cases. In cases where no prior data is available, we need a procedure to optimise the hyperparameter $\alpha$ as we gather data from the SUT and test runs. From other experiences with our client, we believe that $\alpha$ will vary a lot based on the pace of testing and the SUT itself.

In the future, we are planning to apply the sensitivity prioritisation method with other clients in different industrial domains as well as to implement the resource managed regression testing framework.

%\begin{acknowledgement}
%
%\end{acknowledgement}


\begin{thebibliography}{widest}

%STYLE
%\bibitem{haraty}
%{\sc R.A.~Haraty, N.~Mansour, L.~Moukahal, I.~Khalil},\,
%{\sl name of the article},
% Journal, volume(number) (year), pp--pp.

\bibitem{ekelund}
{\sc E.D.~Ekelund, E.~Engstr\"om},\,
{\sl Efficient Regression Testing Based on Test History: An Industrial Evaluation},
IEEE International Conference on Software Maintenance and Evolution (ICSME), 2015, 449--457.

\bibitem{runeson}
{\sc E.~Engstr\"om, P.~Runeson},\,
{\sl A Qualitative Survey of Regression Testing Practices},
International Conference on Product-Focused Software Process Improvement (PROFES), 2010, 3--16.

\bibitem{engstrom}
{\sc E.~Engstr\"om, P.~Runeson, M.~Skoglund},\,
{\sl A Systematic Review on Regression Test Selection Techniques},
Information and Software Technology, 52(1) (2010), 14--30.

\bibitem{ali-mantyla}
{\sc A.~Haghighatkhah, M.~M\"antyl\"a, M.~Oivo},\,
{\sl Test Case Prioritization Using Test Similarities},
International Conference on Product-Focused Software Process Improvement (PROFES), 2018, 243--259.

\bibitem{mantyla}
{\sc A.~Haghighatkhah, M.~M\"antyl\"a, M.~Oivo, P.~Kuvaja},\,
{\sl Test Prioritization in Continuous Integration Environments},
The Journal of Systems and Software, 146 (2018), 80--98.

%\bibitem{haraty}
%{\sc R.A.~Haraty, N.~Mansour, L.~Moukahal, I.~Khalil},\,
%{\sl Regression Test Cases Prioritization Using Clustering and Code Change Relevance},
%International Journal of Software Engineering and Knowledge Engineering, 26(5) (2016), 733--768.
% vol. 26, no. 5

%\bibitem{khati}
%{\sc M.~Khatibsyarbini, M.~A.~Isa, D.~N.~A.~Jawawi, R.~Tumeng},\,
%{\sl Test Case Prioritization Approaches in Regression Testing: A Systematic Literature Review},
%Information and Software Technology, 93 (2018), 74--93.

\bibitem{kaushik}
{\sc N.~Kaushik, M.~Salehie, L.~Tahvildari, S.~Li, M.~Moore},\,
{\sl Dynamic Prioritization in Regression Testing},
IEEE International Conference on Software Testing, Verification and Validation Workshop (ICSTW), 2011, 135--138.

\bibitem{kim}
{\sc J.M.~Kim, A.~Porter},\,
{\sl A History-Based Test Prioritization Technique for Regression Testing in Resource Constrained Environments},
IEEE International Conference on Software Engineering (ICSE), 2002, 119--129.

\bibitem{ledru}
{\sc Y.~Ledru, A.~Petrenko, S.~Boroday, N.~Mandran},\,
{\sl Prioritizing Test Cases with String Distances},
Automated Software Engineering, 19(1) (2012), 65--95.

%\bibitem{osinski}
%{\sc S.~Osi{\'n}ski, J.~Stefanowski, D.~Weiss},\,
%{\sl Lingo: Search Results Clustering Algorithm Based on Singular Value Decomposition},
%Intelligent Information Processing and Web Mining, Springer, 359--368, 2004.

\bibitem{rothermel}
{\sc G.~Rothermel, M.J.~Harrold},\,
{\sl Analyzing Regression Test Selection Techniques},
IEEE Transactions on Software Engineering, 22(8) (1996), 529--551.

\bibitem{sherriff}
{\sc M.~Sherriff, M.~Lake, L.~Williams},\,
{\sl Prioritization of Regression Tests using Singular Value Decomposition with Empirical Change Records},
IEEE International Symposium on Software Reliability Engineering (ISSRE), 2007, 81--90.

\bibitem{yoo-survey}
{\sc S.~Yoo, M.~Harman},\,
{\sl Regression Testing Minimization, Selection and Prioritization: A Survey},
Software Testing, Verification and Reliability, 22(2) (2012), 67--120.

\end{thebibliography}
\end{document}